\documentclass[aps,
amsmath,amssymb,
prb,
superscriptaddress,
reprint,
showpacs
]{revtex4-1}

\usepackage[pdftex]{graphicx}
\usepackage{nicefrac}
\usepackage[normalem]{ulem}
\usepackage{color}

\begin{document}

\title{Making ARPES Measurements on Corrugated Monolayer Crystals: Suspended Exfoliated Single-Crystal Graphene}

\author{Kevin R. Knox}
\affiliation{Department of Physics, Columbia University}
\affiliation{Department of Applied Physics, Columbia University}

\author{Andrea Locatelli}
\affiliation{Elettra - Sincrotrone Trieste S.C.p.A., Basovizza, Trieste, Italy}

\author{Mehmet B. Yilmaz}
\affiliation{Department of Physics, Fatih University, Istanbul, Turkey}

\author{Dean Cvetko} 
\affiliation{CNR-IOM, Laboratorio Nazionale TASC, Trieste, Italy}
\affiliation{Faculty
for Mathematics and Physics, University of Ljubljana, Ljubljana, Slovenia}

\author{Tevfik Onur Mentes}
\affiliation{Elettra - Sincrotrone Trieste S.C.p.A., Basovizza, Trieste, Italy}

\author{Miguel \'Angel Ni\~no}
\affiliation{Elettra - Sincrotrone Trieste S.C.p.A., Basovizza, Trieste, Italy}
\affiliation{Instituto Madrile\~no de Estudios Avanzados IMDEA Nanociencia, Cantoblanco, 28049 Madrid, Spain}

\author{Philip Kim}
\affiliation{Department of Physics, Columbia University}

\author{Alberto Morgante}
\affiliation{CNR-IOM, Laboratorio Nazionale TASC, Trieste, Italy}
\affiliation{Department of Physics, Trieste University, Trieste 34127, Italy}

\author{Richard M. Osgood, Jr.}
\affiliation{Department of Applied Physics, Columbia University}

\date{\today}

\begin{abstract}
Free-standing exfoliated monolayer graphene is an ultra-thin flexible membrane, which exhibits out of plane deformation or corrugation. In this paper, a technique is described to measure the  band structure of such free-standing graphene by angle-resolved photoemission. Our results show that photoelectron coherence is limited by the crystal corrugation. However, by combining surface morphology measurements of the graphene roughness with angle-resolved photoemission, energy-dependent quasiparticle lifetime and bandstructure measurements can be extracted. Our measurements rely on our development of an analytical formulation for relating the crystal corrugation to the photoemission linewidth. Our ARPES measurements show that, despite significant deviation from planarity of the crystal, the electronic structure of exfoliated suspended graphene is nearly that of ideal, undoped graphene; we measure the Dirac point to
be within 25 meV of $E_F$. Further, we show that suspended graphene behaves as a marginal Fermi-liquid, with a quasiparticle lifetime which scales as $(E - E_F )^{-1}$;
comparison with other graphene and graphite data is discussed.

\end{abstract}


\maketitle

\section{\label{sec:level1}Introduction}

The recent availability of monolayer-thick two-dimensional crystals such as graphene, BN, and BSCCO has generated widespread interest in the physics
and materials science communities. In the case of graphene, in particular, the two dimensional nature of the crystal in combination
with its unusual massless Dirac fermions determines a host of intriguing and unique transport phenomena, including graphene's half-integer quantum
Hall effect (HE) and non-zero Berry's phase.\cite{novoselov2005two,zhang2005experimental}
Unlike most metals, undoped graphene has a Fermi surface which consists of a set of 2 inequivalent points in momentum-space.
Thus, at zero temperature and zero doping, the density of states at the Fermi
level vanishes. In combination with the linear dispersion of low energy charge
carriers, this vanishing density of states is expected to lead to unusual
band-renormalization effects that are not seen in Fermi-liquid systems such
as unusually high electron-electron coupling. Motivated by interest
in these unusual properties, several theoretical and experimental studies have investigated the electronic properties of graphene.\cite{neto2009electronic}

Angle-resolved photoemission spectroscopy (ARPES) is the experimental method that is most frequently used to probe the electronic structure of crystals. However, so far, the majority of ARPES studies of graphene have been conducted on epitaxial graphene, which has been grown on a variety of substrates such as SiC, Ru, Ni and Ir. \cite{ohta2006controlling, bostwick2006quasiparticle, dedkov2008rashba, shikin2000surface, vazquez2008periodically, liu2010phonon, pletikosic2009dirac,sutter2009electronic} Epitaxial graphene is ideal for photoemission experiments, but, due to the interaction between the epitaxial graphene monolayer and the substrate, the band structure is often distorted
such that the Dirac point shifts away from the Fermi energy, thus changing
the quasiparticle dynamics.  In an effort to minimize the effect of substrate interaction on epitaxial graphene, recent ARPES studies have focused on several multilayer systems, such as intercalated graphite\cite{valla2009anisotropic} and graphene grown on the C face of SiC. \cite{sprinkle2009first} These layered systems consist of multiple stacked graphene sheets that are substantially electrically isolated, thus resulting in an electronic band structure that mimics that of suspended exfoliated single-layer graphene. However, despite its scientific and technological importance, exfoliated graphene has been the subject of only a limited number of ARPES studies,\cite{knox2008spectromicroscopy,liu2010phonon}
despite the fact that it remains the best choice for device physics, as it
is easily backgated and has the highest
measured mobility.\cite{bolotin2008ultrahigh}

Several obstacles impede measurement of the bandstructure
of exfoliated graphene. One difficultly arises from the fact that available
single-layer exfoliated graphene flakes are typically less than 20 $\mu$m in size,
thus precluding the use of standard ARPES
systems, which require samples to be several mm in size.  Hence, most information
regarding low-energy occupied states in exfoliated graphene has been obtained
indirectly from electrical-transport measurements\cite{novoselov2005two,zhang2005experimental}
or directly by optical-probing techniques.\cite{wang2008gate,mak2010evolution} These
techniques examine the bandstructure generally within ~1eV of the Dirac point
and do not directly provide momentum resolution.  For photoemission the limitation in size
can
be overcome by working with high lateral-spatial-resolution probes such as
those available using spectromicroscopy.\cite{fujikawa2009micrometer,sutter2009electronic} A second major impediment to photoemission studies is due to the fact that graphene is an ultrathin crystal.
This ultrathin property has, in turn, two important
consequences for photoemission studies. The first is the transparency of monolayer graphene to UV photons and photoemitted electrons, which causes a strong background photoemission signal if the monolayer
graphene is in close physical proximity with a substrate.\cite{knox2008spectromicroscopy}
The second is that exfoliated graphene is not atomically
flat, but is known to deform
locally, a result shown through AFM, STM, electron
microscopy, and electron scattering results.\cite{geringer2009intrinsic,
ishigami2007atomic,locatelli2010corrugation,stolyarova2007high,cullen2010high}
It has been argued that the deformation is due to the fact that monolayer-thick graphene has soft flexural
modes leading to ready bending of the graphene.  The presence of a supporting
substrate or scaffold can, to a certain degree, stabilize height fluctuations in the graphene layer, but corrugations in the underlying supporting substrate are transferred in part to the graphene due to the reduced stiffness of this material. Additionally, intrinsic corrugations that cannot be attributed to interaction
with the substrate were recently observed in supported graphene.\cite{geringer2009intrinsic} Further, in a recent low energy diffraction study, we demonstrated that even graphene suspended over etched cavities exhibits corrugation, which appeared to have been intrinsic in origin.
\cite{locatelli2010corrugation}

Thus, in general, two dimensional crystals produced by exfoliation may
show significant local curvature, manifested as corrugation and ripples. This corrugation is known to affect not only the electronic and transport properties of the material, but can also have
a major impact on photoemission results. In particular, the theory of
ARPES was developed for single-crystal atomically flat surfaces and relies on the fact that momentum perpendicular to the surface is conserved in the photoemission process. On such perfectly ordered crystals the photoemission
lineshape is directly related to the spectral function of the electronic
state being probed, from which information about many-body physics can
be extracted. 
The corrugation in thin sheets of layered materials breaks this symmetry and obscures the intrinsic many-body effects.  

In this paper, we present a systematic approach to account for such corrugation-induced broadening in ARPES on thin films. By combining our photoemission
results with detailed information
about surface morphology obtained from prior electron-microscopy measurements\cite{locatelli2010corrugation}
taken in-situ on the same samples
we are able to quantify the influence of corrugation on spectral broadening.
We go on to describe a method to discount the effect of surface corrugation from ARPES measurements to reveal the intrinsic many-body physics present
in graphene. Our results show that suspended graphene behaves as a marginal Fermi-liquid with an anomalous quasiparticle lifetime which scales as $(E - E_F )^{-1}$.

\section{\label{sec:level2}Experiment}

Our measurements used the Spectroscopic Photoemission and Low Energy Electron
Microscope (SPELEEM) at the Nanospectroscopy beamline at the Elettra Synchrotron
light source.\cite{nano01} The SPELEEM is a versatile multi-technique microscope that combines low energy electron microscopy (LEEM) with energy-filtered X-ray photoemission electron microscopy (XPEEM). The microscope images surfaces, interfaces and ultra-thin films using a range of complementary analytical characterization methods, which have been described in detail previously. \cite{schmidtetalSRL98, locatelliandbauerJPCM2008} When operated as a LEEM, the microscope probes the specimen using elastically backscattered electrons. LEEM enables high sensitivity to surface crystalline structure and, due to the favorable backscattering cross-sections of most materials at low energies, allows image acquisition to be obtained at video frame rate. The lateral resolution of the microscope for LEEM imaging is currently below 10 nm. In XPEEM mode, the specimen is probed using the beamline photons, provided by an undulator source; thus,
the technique is sensitive to the local chemical and electronic structures. Laterally resolved versions of synchrotron based absorption (XAS) and photoemission spectroscopy (XPS) are possible. The lateral resolution in XPEEM approaches a few tens of nm.\cite{locatelli2006photoemission}

Along with real-space imaging, the SPELEEM microscope is capable of micro-probe diffraction imaging, i.e.\ laterally restricted low energy electron diffraction ($\mu-$LEED) and angle resolved photoemission electron spectroscopy ($\mu-$ARPES) measurements when probing with electrons and photons, respectively. In diffraction operation the microscope images and magnifies the back focal plane of the objective lens. In ARPES mode, the full angular emission pattern can be imaged on the detector up to a parallel momentum of $\sim 2 \text{\AA}^{-1}$; at larger parallel momentum the transmission of the microscope decreases. All diffraction measurements are restricted to areas of $\sim$ 2 $\mu m$ in diameter, which are selected by inserting a field limiting aperture into the first image plane along the imaging-optics column of the instrument.  Thus, the microscope enables measurements on samples that are homogeneous over areas of a few square microns. The energy resolution of the SPELEEM in diffraction imaging, such as ARPES and photoelectron-diffraction measurements, is 300 meV and the transfer width of the microscope when operated in LEED mode
is 10 nm.

\begin{figure}
\includegraphics[width=8cm]{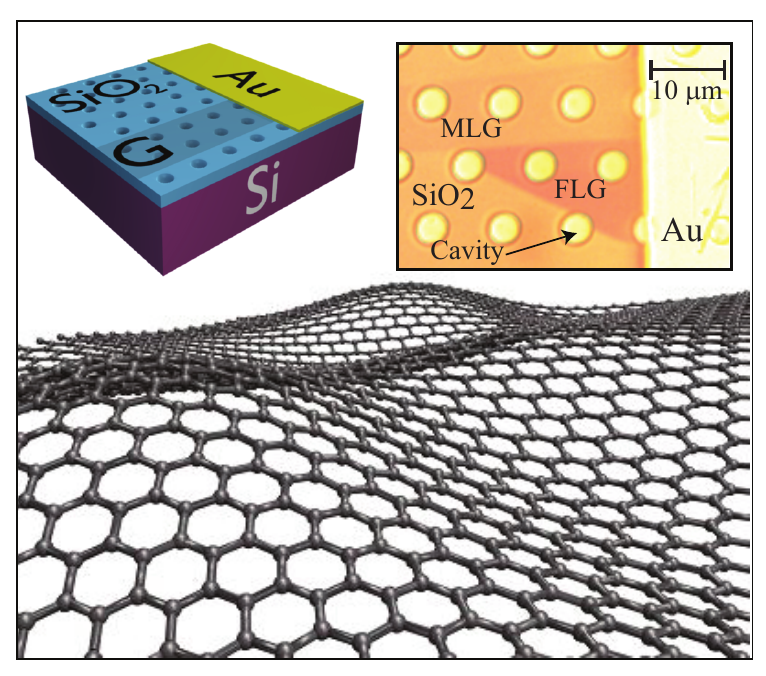}
\caption{\label{fig:samplePic} (Color Online) (Left) Schematic drawing of our suspended-graphene
sample configuration. (Right) Optical micrograph of sample containing suspended
monolayer graphene (MLG) and few-layer graphene (FLG). (Bottom) Artists rendering
of corrugated graphene crystal (height fluctuations not to scale).}
\end{figure}

Graphene samples were extracted by micro-mechanical cleavage from Kish graphite
crystals (Toshiba Ceramics, Inc.) and placed onto an SiO$_2$-thin-film layer
on an Si substrate, which was previously patterned with cylindrical cavities
to a depth of 300 nm, as described in Ref. \onlinecite{locatelli2010corrugation}. The use of planar processing
of this substrate allowed us to suspend areas of the graphene films without
the use of further photolithographic techniques, which would introduce contaminants
to the graphene sheets. Graphene samples with lateral sizes from
10 to 50 $\mu$m were placed in contact with Au grounding strips deposited on the
surface via thermal
evaporation through a shadow mask. A sketch of the sample configuration
is shown in Fig. 1 along with an optical micrograph.  

The SPELEEM instrument used to collect data has the important advantage of
having a sufficiently high spatial resolution to guarantee that we are measuring
a single crystal sample of monolayer graphene and that all of the measured
spectral intensity is derived from a fully suspended region. This capability
is necessary since the suspended regions are approximately 5 $\mu$m in diameter
and, therefore, cannot be resolved with conventional photoemission instruments,
which employ spatial averaging techniques that collect data over surface
areas of several square millimeters. The potential to combine both photoemission and electron scattering measurements is essential for our experiment
since it allows us to measure bandstructure
and surface morphology on the same samples. We note, additionally, that a
similar instrument was recently
used in a study, which examined the morphology and electronic structure of epitaxial graphene grown on Pt.\cite{sutter2009electronic}

After preparation the samples were placed into a UHV chamber with a base
pressure of $2 \times 10^{-10}$ mbar, and the surface cleaned via low energy electron irradiation to eliminate adventitious hydrocarbon molecules
adsorbed during prior atmospheric exposure.\cite{locatelli2010corrugation}  All graphene samples were characterized with LEEM before investigation
with ARPES and LEED.  
For each sample, LEEM was used to locate sample areas of interest and to determine film thickness with atomic resolution by measuring intensity modulations in the
LEEM I-V spectra.\cite{locatelli2010corrugation,hibino2008microscopic,altman2005low}

ARPES data at multiple
photon energies were obtained on the suspended areas of the graphene film.
Only regions of uniform thickness were considered. In order to elucidate the role of surface corrugation and substrate influence, comparative experiments were also carried out on corresponding regions where the film was supported by the SiO$_2$ substrate. 
This surface has been recently carefully
calibrated by prior STM and electron-scattering measurements.
\cite{ishigami2007atomic,stolyarova2007high,geringer2009intrinsic}
In addition, ARPES measurements were made
on Kish-graphite flakes that were present on the same substrates. As graphite is a well understood and commonly studied
system, these measurements provided a useful point of comparison for our
graphene measurements. Photoemission from graphite is, in some respects, similar to that from graphene
because of the stacked-layer nature of the former.  However, the physics
near the Dirac point is significantly different owing to the fact that the
multilayer stacking in graphite breaks the symmetry between A and B sublattices,
which results in two dispersing branches, such that low energy excitations
do not have the simple linear dispersion relation that is found for graphene.

\section{\label{sec:Results}Results}

\begin{figure}
\includegraphics[width=8cm]{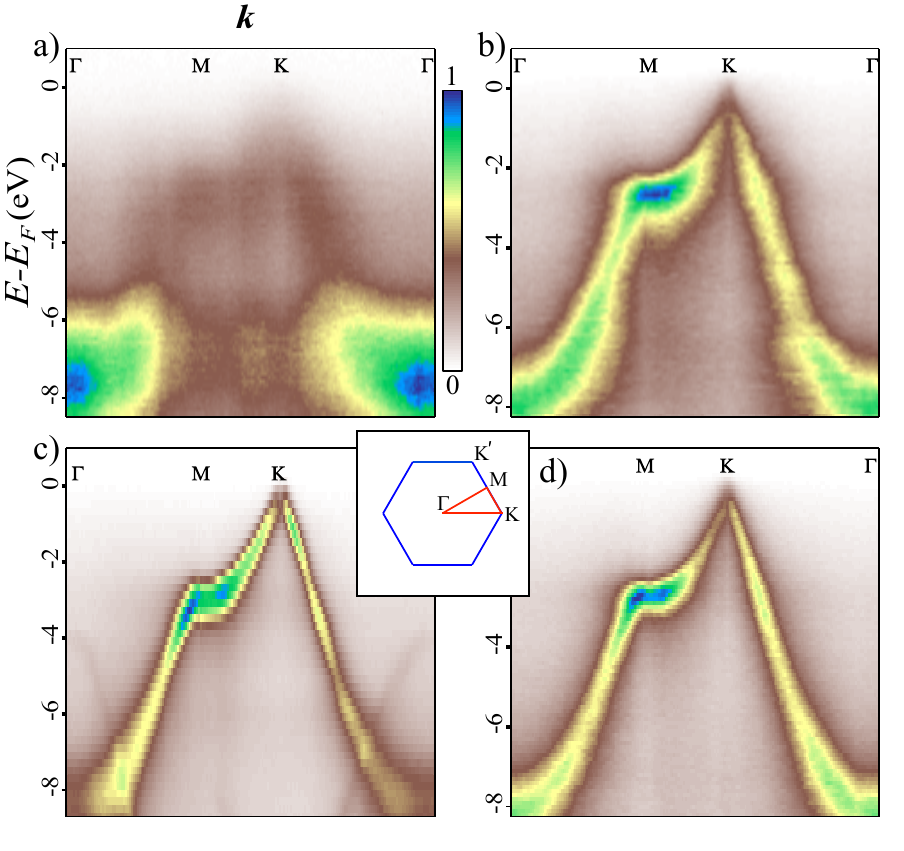}
\caption{\label{fig:rawARPES} (Color online) Raw ARPES data along symmetry directions
in Brillouin Zone for graphene and graphite. (a)SiO$_2$ supported graphene
($\hbar \omega$=90 eV). (b) Suspended graphene ($\hbar \omega$=84 eV). (c)Kish
graphite ($\hbar \omega$=90 eV). (d) Suspended graphene ($\hbar \omega$=50 eV). Inset shows 2D graphene Brillouin zone.}
\end{figure}

Photoemission spectra were measured from two samples with
differing degrees of surface corrugation and substrate interaction, that
is, on suspended and substrate-supported graphene. Previous LEED measurements have shown that the horizontal correlation length increases from 24 nm to 30 nm 
in measurements taken on supported and suspended
samples, respectively.\cite{locatelli2010corrugation} In addition, ARPES data were collected at room temperature over the entire surface Brillouin zone (SBZ) from 0.5 eV to -8 eV (energy referenced to $E_F$), for monolayer graphene and
graphite, using a range of photon energies.  Figure \ref{fig:rawARPES}
shows ARPES spectra taken from a sample supported by and in contact with
the SiO$_2$ surface and a sample that was suspended over the 5 $\mu$m wells
shown in Fig. \ref{fig:samplePic}. For comparison,
the raw ARPES spectrum from Kish graphite is shown as well. The data show
dispersion along 3 symmetry lines in the SBZ. As expected from the reduced corrugation, as
well as the absence of any substrate interaction, the ARPES data for suspended
graphene show a dramatic improvement in quality as compared to the data
for supported graphene.  Additionally,
there is a very broad, parabolically dispersing peak centered at the $\Gamma$
point at a binding energy of $\sim$8 eV in the data taken on \emph{supported} graphene.
This feature has been previously attributed to photoemission from the amorphous
SiO$_2$ substrate\cite{knox2008spectromicroscopy} and is not observed in the spectrum taken on suspended
graphene. Although the substrate is only 300 nm below the suspended graphene,
any background electrons emmited at this height will be significantly defocused
in the electron optics of SPELEEM microscope. Additionally, due to the grazing
incidence angle of the photon beam (16$^{\circ}$), the bottom of the cavity
is not fully illuminated as the cavity edge casts a shadow, which further
reduces the photoemission signal from the substrate.

\begin{figure}
\includegraphics[width=8cm]{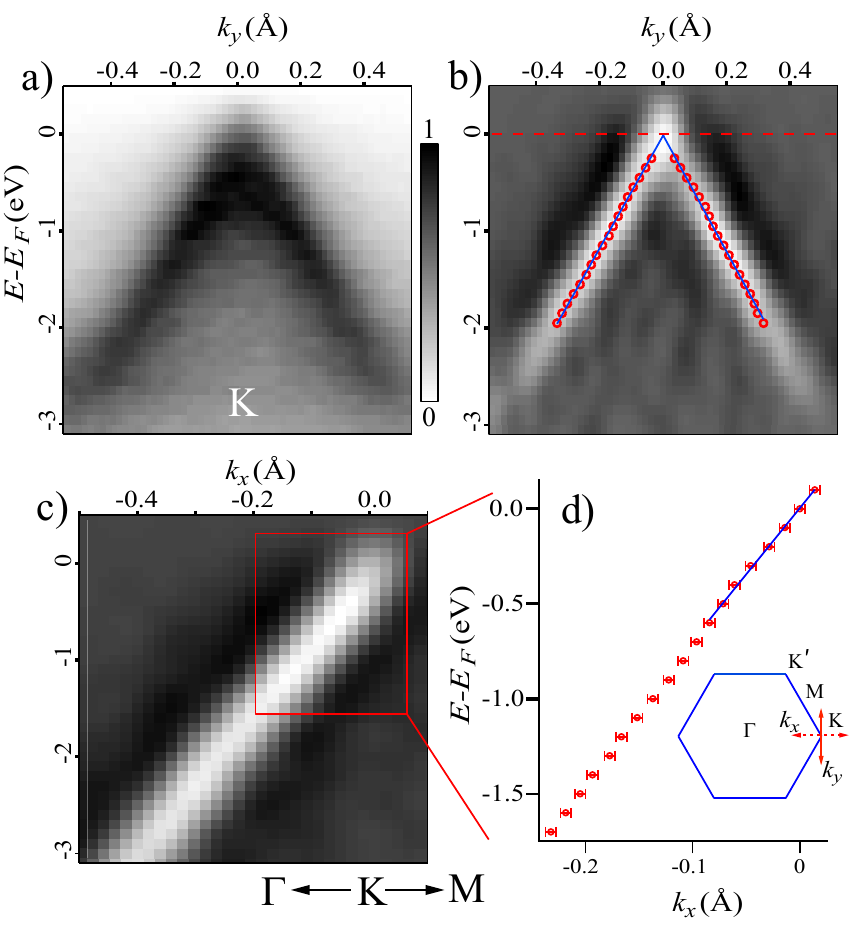}
\caption{\label{fig:grapheneARPES}  (Color online) (a) ARPES intensity through K point along $\Gamma$M ($k_y$) direction in suspended monolayer graphene ($\hbar \omega = 50 eV$). (b) Smoothed second derivative image of dispersion shown in (a).  (c) Smoothed
second derivative ARPES
intensity through K point along $\Gamma$K ($k_x$) direction.  (d) Extracted dispersion
from (c) Inset shows graphene Brillouin zone. Red solid (dashed) line indicates
$k_y$ ($k_x$) direction through K point.
}
\end{figure}

In the vicinity of the K points, a conical dispersion is observed centered
at the K point on the suspended graphene spectrum.  At
$\sim 1$ eV below
the Fermi level a trigonal-warping deviation from angular isotropy becomes
clearly noticeable. Measurements taken through the K point and in the direction parallel to the
$\Gamma$M direction (vertical direction) show two symmetric
dispersing branches forming the two sides of the Dirac cone.
The band structure can be made significantly sharper (see Fig. \ref{fig:grapheneARPES})
by taking the second derivative along each momentum direction. In this case,
use of the second derivative allows easier determination of the Dirac point
with respect to the Fermi level. Figure \ref{fig:grapheneARPES}(b) shows the
linear best fit to the two branches as well as the location of the Fermi
level. From the fit, we find that the Dirac point is within 25 meV of $E_F$ ($ E_D = -  9 \pm 25 meV)$. Thus,
the sample is minimally doped due to the preparation
procedure used here, which did not involve any photolithographic or chemical-transfer techniques. In contrast, the Dirac point previously measured by
our group on a supported sample was found to be $\sim$300 meV below  the
Fermi level, which was attributed to doping by interaction with charged impurities
in the SiO$_2$ layer.\cite{knox2008spectromicroscopy}

\begin{figure}
\includegraphics[width=8cm]{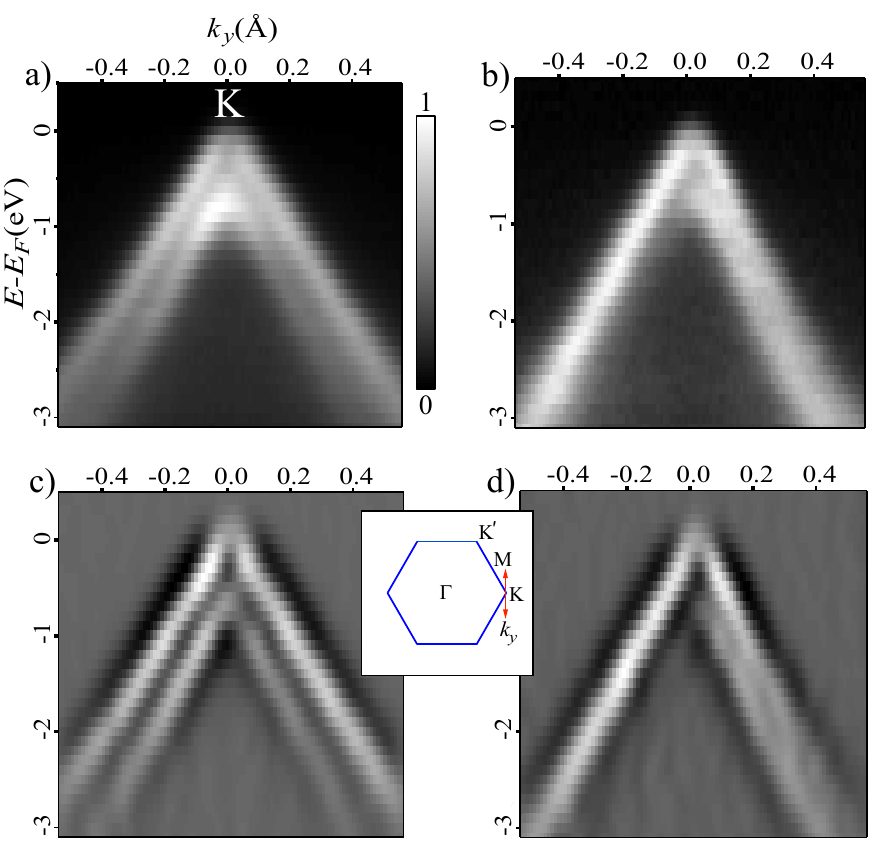}
\caption{\label{fig:graphiteARPES} (Color online) Dispersion along the vertical ($k_y$) direction
through the \=K point for graphite obtained at photon energies of (a) $\hbar
\omega = 86$ eV and (b) $\hbar \omega = 76$ eV. (c-d) Smoothed second derivative images
of spectra shown in (a) and (b), respectively. Inset shows graphite surface
Brillouin zone. Solid red line indicates $k_y$ direction through \=K point.}
\end{figure}

For comparison with results on a known photoemission materials system, graphite
spectra were taken at two photon energies (86 and 76 eV) along the same (vertical)
direction through the K point; these results are shown in Fig. \ref{fig:graphiteARPES}.
The dispersion obtained at $\hbar \omega$=86 eV is clearly symmetric about
the K point.  At this photon energy we can resolve the splitting of the $\pi$
state into bonding and antibonding bands, with the two bands separated by $\sim$ 0.12 $\text{\AA}^{-1}$.
The bands themselves are approximately 0.1 $\text{\AA}^{-1}$ in width.
In the spectrum taken at 76 eV the two peaks are nearly degenerate. Again, the
second derivative allows for easier determination of peak locations.

\begin{figure}
\includegraphics[width=8cm]{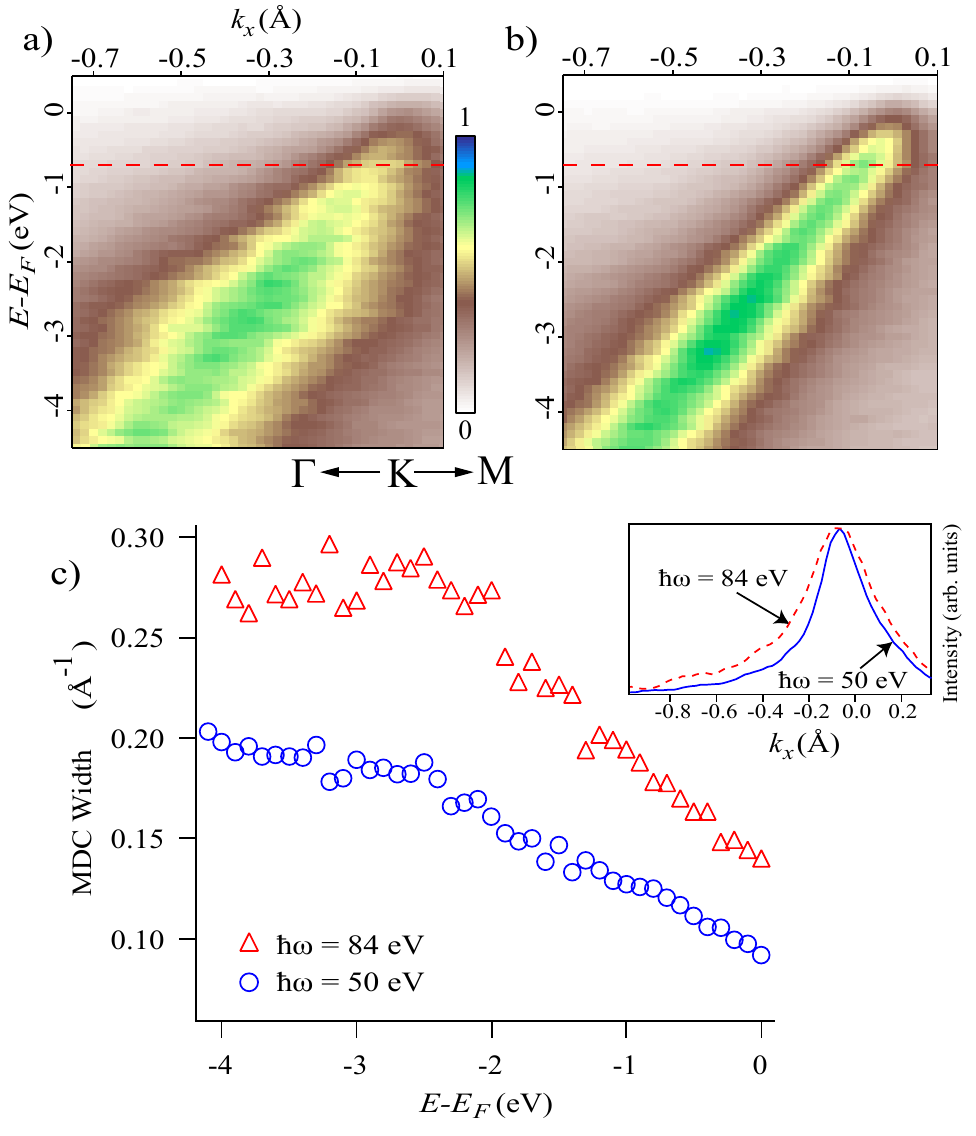}
\caption{\label{fig:grapheneKpoint} (Color online) ARPES intensity along $\Gamma$K direction in suspended monolayer graphene. Graphene photoemission
taken with photon energies of (a) $\hbar \omega = 50 eV$ and (b) $\hbar \omega = 84 eV$. (c) HWHM of MDCs as a function of binding energy taken from (a) and (b). Inset shows sample MDCs
taken 0.7 eV below $E_F$ as indicated by dashed red lines in (a) and (b).
}
\end{figure}

Figure \ref{fig:grapheneKpoint} shows the graphene dispersion taken along
the $\Gamma$K  direction through the K point.
Comparative measurements were made at two different photon energies ($\hbar
\omega =50,\hbar \omega =84$) and are shown in Fig. \ref{fig:grapheneKpoint}(a)
and (b), respectively. In this direction, only one branch of the dispersion
can be seen as the difference in phase between electron waves emitted from
the A and B sub-lattice sites results in complete destructive interference.
\cite{shirley1995brillouin} Thus, this is a convenient direction along which to measure precisely
the dispersion in the vicinity of the Dirac cone.  The inset to Figure
\ref{fig:grapheneKpoint}(c) compares momentum distribution curves (MDCs) taken
at a binding energy of 0.7 eV for the suspended-graphene spectra at both
photon energies.
The data in Fig. \ref{fig:grapheneKpoint}(c) show that the width of the $\hbar \omega$=84 eV MDC is significantly larger than the
$\hbar \omega$=50 eV MDC (0.17 $\text{\AA}^{-1}$ vs 0.12 $\text{\AA}^{-1}$).

Additionally, there is a slight asymmetry in all three MDCs as additional
spectral weight is present on the right side of the peak (at higher values
of $k_x$). The background signal decreases and the peaks become narrower
for 50 eV photons as compared to 84 eV photons. Specifically, in the 0-4
eV range (referenced to $E_F$), the MDC width increases monotonically from
0.1 to 0.2 $\text{\AA}^{-1}$ and from 0.15 to 0.3 $\text{\AA}^{-1}$ for data
collected with 50 eV photons and 84 eV photons, respectively. In contrast,
MDCs taken along the same direction ($\Gamma$K) on supported graphene are
significantly broader\cite{knox2008spectromicroscopy} and show almost no
dependence on binding energy; they are $\sim$0.5 $\text{\AA}^{-1}$ in width
from the Fermi energy to -4 eV binding energy. Thus, spectral features are sharpest for suspended
samples measured with lower photon energy.

\section{\label{sec:Discussion}Discussion}
   
 \subsection{\label{sec:GraphiteVsGraphene}Comparison of Graphite and Graphene
 Results}

As shown in Fig. \ref{fig:grapheneKpoint}, variation in photon energy results in changes to
the linewidth of the graphene photoemission spectra, which can be exploited to
sharpen the spectrum.  In explaining these results on graphene, it is useful first to examine the effect of photon energy variation for the case of \emph{graphite}.
The differences in the measured photemission spectra of graphite taken at
$\hbar \omega$=76 eV and $\hbar \omega$=86 eV shown in Figs. \ref{fig:graphiteARPES}(a) and (b), respectively, are easily understood by considering the 3 dimensional band structure of
graphite. In particular, according to the standard model of photoemission,
variation in 
$\hbar \omega$ allows one to access
a range of initial states with different $k_z$.\cite{Hufner} Using the free-electron
approximation for the final state allows calculation of $k_z$ of the initial
state.\cite{Hufner,law1986synchrotron} Thus, in the case of the data shown
in Fig.
\ref{fig:graphiteARPES},
photoemission obtained at $\hbar \omega$ = 86 eV corresponds
to $k_z$ = 0 (the $\Gamma$KM plane), while $\hbar \omega$ = 76 eV accesses $k_z$ = 0.3$c^{\ast}$ which is nearer the AHL plane.
Since $k_z$ is changed, the clear double band feature seen for $k_z=0$ changes as the graphite band structure varies along $k_z$ in accord with the known graphite band structure.
\cite{shirley1995brillouin,zhou2006first,sugawara2007anomalous,law1986synchrotron}
   
Consider now the effect of changing photon energies for the case of graphene
photoemission. Since graphene is truly a 2D crystal,
the initial states in the valence band are highly localized along the $z$
direction. Thus, the Brillouin zone is strictly 2 dimensional and the electronic strucure is essentially $k_z$ independent. Comparison
with photoemission from surface states is useful, since they
are also localized in 2D.\cite{kevan1987high}
However, the role of evanescent decay into the bulk, which is important for
surface states in metals and results in a partial $k_z$ dependence,\cite{kevan1987high}
such as surface resonance, is absent in graphene and, thus, we may treat the initial state as independent of photon energy. In fact, as seen in Fig. \ref{fig:grapheneKpoint}, changing the photon energy in the case of graphene causes only a change in
the overall linewidth and does not affect the measured bandstructure.
As will be discussed below, the difference in the width of ARPES features
between
spectra obtained at $\hbar \omega$ =50 and $\hbar \omega$ =84 is a consequence
of the surface roughness of the graphene samples. Since electrons in graphene
propagate on a locally curved surface, the usual momentum conservation rules
in ARPES must be modified and a photon-energy-dependent broadening
term is introduced.

\subsection{\label{sec:GeneralConsiderations}General Considerations}
        
In standard many-body ARPES theory, the intensity of the photoemission signal is proportional to the spectral function, $A({\bf k},\omega)$:
        
\begin{equation}
A({\bf k},\omega)=\frac{\text{Im} [\Sigma({\bf k},\omega)]}{(\omega-\omega_{\bf k}-\text{Re} [\Sigma({\bf k},\omega)]
)^2+\text{Im} [(\Sigma({\bf k},\omega)]^2}
\end{equation}
where $\omega=E-E_F$ and ${\bf k}$ are binding energy and momentum, respectively, and $\omega_{\bf k}$ is the single-particle dispersion. The real and imaginary parts of the self-energy, $\Sigma({\bf k},\omega)$, represent renormalization of the bare-bands and scattering rate, respectively. To obtain the full expression
for the photocurrent, the above function is then multiplied by energy and
momentum-preserving delta functions, $\delta({\bf k}_i -{\bf k}_f-{\bf G})
\delta(E_i-E_f-W)$, where $\bf G$ is a reciprocal lattice vector and i and
f label the initial and final states, respectively, and $W$ is the work function
of the material.  

However, one major complication to this approach arises for the case of suspended
graphene since the momentum preserving function, $\delta({\bf k}_i -{\bf k}_f-{\bf G})$, is only a precise delta-function if the system under
investigation is atomically flat. While this is the case for the majority
of single-crystal samples probed with ARPES, including the Kish graphite
described above, exfoliated
monolayer graphene, as is discussed in the Introduction, has significant deviations from planarity,
ranging from 1 to 10 $\text{\AA}$.\cite{meyer2007structure} This corrugation introduces
an additional broadening mechanism into the ARPES spectrum, which can be as
large as, or larger than, the intrinsic broadening represented by $\text{Im} [\Sigma({\bf k},\omega)]$. Thus, in order
to extract the true self-energy of carriers in the crystal, such corrugation-induced broadening must be taken into account. The
MDCs are best fit by a convolution of $A({\bf k},\omega)$ with a function
that represents broadening due to surface roughness. Thus, as will be shown
below, at fixed $\omega$, photoemission intensity
as a function of ${\bf k}_\|$ can be expressed as:
 
\begin{equation}
  I({\bf k}_\|) \propto \int d^2{\text k_\|}' S_{k_\perp}({\bf k}_\|) 
  A({\bf k}_\|-{\bf k}'_\|,\omega)
\label{eq:convBroad}
\end{equation}
where ${\bf k}_{\|}={\bf k}_{i\|}-{\bf k}_{f\|}$ and $S_{{\bf k}_\perp}$ represents corrugation-induced broadening. $S_{{\bf k}_\perp}$, the surface
structure factor, is a function of
the surface geometry of the sample and is also generally dependent
on the change in perpendicular momentum from initial to final state, $k_{\perp}=k_{i\perp}-k_{f\perp}$.
We note that several prior studies have examined the effect of surface roughness
on ARPES measurements.\cite{theilmann1999high,theilmann1997influence} 
In these prior studies, the
roughness considered was due to discrete height variations caused by monatomic steps, rather than
the continuous undulations of a thin film. Thus, the broadening in spectral features
measured by ARPES was attributed to increased electron scattering rather
than a variation in the phase of photoemitted electrons induced by local height fluctuations.
In our experiments on suspended graphene samples, the surface 
morphology
is carefully measured simultaneously with
the ARPES measurements presented here, thus allowing us
to determine $S_{{\bf k}_\perp}$ independently.\cite{locatelli2006photoemission}

Finally, note that the surface corrugation of the graphene sheets will also alter
the bandstructure by inducing a change in the local potential proportional to the square of the curvature. Thus, the ripples act as scattering centers, which will decrease lifetime and potentially change the Fermi velocity. These effects are contained in $ A({\bf k},\omega)$ and will also be present in the ARPES data.
However, such effects are distinct from that described by $S_{k_\perp}$, which represents decoherence as electrons pass from a curved 2D space to free space.

\subsection{\label{sec:CorrugationBroadening}Corrugation Broadening}

Corrugation broadening can be treated by considering the equation that describes
photoemission from a Bloch state in the graphene sheet into a free-electron
state above the crystal. Using the standard tight-binding approach to describe
the initial state: 

\begin{equation}
  \psi_{\bf k} ({\bf r})=\frac{1}{\sqrt{N}} \sum_{\bf R} e^{i {\bf k} \cdot
  {\bf R}} \sum_{j=A,B} C^{\bf k}_j \phi ({\bf r}-{\bf R}-\tau_j)  
\label{eq:initState}
\end{equation}
we obtain the following matrix element for excitation into a free electron
final state:

\begin{equation}
  M \propto ({\bf k_i} \cdot \hat{\bf{\lambda}}) \sum_{j=A,B} C^{\bf
  k_i}_j e^{-i{\bf k_f} \cdot \tau_j}
  \sum_R e^{i ({\bf k_i}-{\bf k_f}) \cdot {\bf R}} \tilde{\phi}({\bf k_f})
\label{eq:transitionM}
\end{equation}
where ${\bf k_i}$ is the initial pseudo-momentum of a valence-band electron and ${\bf k_f}$ is the
final-state momentum (for a full description and definitions of symbols see Appendix). Equation \ref{eq:initState} describes an initial state
with precise momentum at a fixed binding energy. 
For an atomically-flat crystalline 2D surface the position vectors
can be expressed as ${\bf R} = n_1 \textbf{a}_1+ n_2 \textbf{a}_2$, where
the
$n_i$ are integers and 
the $\textbf{a}_i$ primitive lattice vectors in the $xy$ plane. In this case,
the sum over
{\bf R} in
Eq. \ref{eq:initState}, $\sum e^{i ({\bf k_i}-{\bf k_f}) \cdot {\bf R}}$,
 is zero unless ${\bf k_{i\|}}-{\bf k_{f\|}} = {\bf G}$, where ${\bf G}$ is
 a reciprocal
lattice vector. This condition is, thus, a statement of the momentum conservation
discussed above,  $\delta({\bf k_i} -{\bf k_f}-{\bf G})$. If, however, $z$ is allowed to vary continuously as a function of position
along the surface, so that ${\bf R} = n_1 \textbf{a}_1+ n_2 \textbf{a}_2 + \Delta x + \Delta y + z$\footnote{
It is implied here that $\Delta x$,  $\Delta y$ and $z$ are functions
of position along the surface with $z$ representing the local height of the
surface and $\Delta x$, $\Delta y$ representing deviations from the ideal
lateral positions of surface atoms which are necessary to keep the average bond length unchanged.},
with $z$ no longer constant, the summation in Eq. \ref{eq:transitionM} is
not as readily calculated. Perfect phase cancellation away from reciprocal
lattice vectors does not occur, resulting in non-zero photoemisison intensity
when ${\bf k_{i\|}}-{\bf k_{f\|}} \neq {\bf G}$.

Summations such as the one in Eq. \ref{eq:transitionM} are encountered
in the theory of LEED on rough surfaces.\cite{lu1982diffraction,yang1992time,yang1993diffraction}
In fact, in many respects the formal analysis of LEED results bears many
similarities to that of ARPES. In a prior
study using one-photon photoemission and high-resolution LEED applied simultaneously to surface states of Cu(100) and Cu(111), it was demonstrated experimentally that the photoemission linewidth and the width of the LEED-spot profile are correlated linearly.\cite{theilmann1997influence}
In particular, for LEED one measures the diffraction structure factor, $S(\bf{k})
\propto | \sum e^{i \bf{k} \cdot {\bf R}} |^2 $ where, as in the case of
photoemission,
$\bf{k}$ is the
total momentum transfer, ${\bf k}={\bf k_i}-{\bf k_f}$, and the sum is over atomic
positions, $\bf R$, on a surface.  In addition, for ARPES transition probability
is proportional to the square of the matrix element; thus, the same
structure factor, $S({\bf k})$, is applicable. Thus, LEED theory can guide
our analysis.  

The structure factor, $S({\bf k})$, can be calculated with information about
the average properties of the surface, described by three variables: horizontal
correlation-length, $\xi$, RMS height variation, $w$, and a dimensionless parameter, $\alpha$, termed the ``roughness exponent," which describes surface roughness on length scales smaller than $\xi$.\cite{yang1993diffraction}
All three parameters can be extracted from real-space information about the
surface by computing the height-height correlation
function, which is used in a variety of thin film measurements, including
those on graphene and other surfaces, and is defined as 
$H(r)=\langle |z(r_0+r)-z(r_0)|^2 \rangle$.
As is shown in the appendix, 
$S({\bf k})=S(k_\perp, {\bf k_\|})$ is intimately related
to $H(r)$ as the Fourier transform of $e^{-\frac{1}{2}k_\bot H(r)}$. 
Thus, the average parameters that characterize a given rough surface
($w$, $\xi$, and $\alpha$) and determine the form of $H(r)$ also determine
$S({k_\perp},{\bf k_\|})$. Hence,
with these parameters, it is possible to compute the summation in Eq. \ref{eq:transitionM}.
In fact, previously reported measurements using low-energy electron microscopy and low-energy
electron diffraction have determined these parameters to be  $\alpha=0.54\pm
0.02$, $w=1.99\pm 0.15$ $\text{\AA}$, and $\xi=30 \pm 0.3$ nm for the same
suspended
graphene samples used in this study.\cite{locatelli2010corrugation} Although
the functional form of $S({k_\perp},{\bf k_\|})$ is complex,
the width of $S_{k_\perp}({\bf k_\|})$ (i.e.\ for $k_\perp$ fixed) in ${\bf k_\|}$
space
has a simple dependence on $k_\perp$ and the parameters describing the surface
roughness. In particular, the width,  
$\Gamma_S$ is proportional to $(k_\bot w)^{\nicefrac{1}{\alpha}}$/$\xi$, which explains the decrease
in experimental linewidth with decreasing $k_\bot$ shown in Fig. \ref{fig:grapheneKpoint}.
For fitting purposes it is useful to have the exact functional form of $S_{k_\perp}({\bf k_\|})$.  Yang, \emph{et al.} have shown that for $(w k_\perp)^2 \gg 1$ the form
is purely diffusive and can be expressed as:\cite{yang1993diffraction}

\begin{eqnarray}
  S_{k_\perp}({\bf k_\|}) = (\xi / (w k_\perp)^{\nicefrac{1}{\alpha}})
  F_{\alpha}({\bf k_\|} \xi / (w k_\perp)^{\nicefrac{1}{\alpha}})
  \nonumber\\
  F_{\alpha}(Y)=\int X dX exp(-X^{2 \alpha} ) J_0 (XY)
\end{eqnarray}

\subsection{\label{sec:IntrinsicBroadening}Intrinsic Broadening}

It is straightforward to introduce intrinsic initial-state broadening into our ARPES description
by 
replacing our initial state wavefunction, $\psi_{\bf k}$, with a sum
over multiple momentum states, $\sum a_{\bf k} \psi_{\bf k}$, where the $a_{\bf k_i}$ are complex
coefficients related to the spectral function by $|a_{\bf k_i}|^2= A({\bf k},\omega)$. 
Our transition matrix then becomes a sum, $M=\sum a_{\bf k} M^{\bf
k}$, over multiple
matrix elements weighted by the complex coefficients $a_{\bf k}$, where the $M^{{\bf k}}$
are the original transition matrix elements defined in Eq. \ref{eq:transitionM}.
Again, using Fermi's golden rule we find that the transition probability
is proportional to the square of this sum.

\begin{equation}
  I \propto |M|^2 = |\sum_k a_k M^{\bf k}|^2 = 
  \sum_{\bf k} |a_{\bf k}|^2 |M^{k}|^2 + \sum_{{\bf k}
  \neq {\bf k}'}a_{\bf k}^* a_{\bf k'} M^{{\bf k}*} M^{\bf k'}
  \label{eq:mBroad}
\end{equation}

As shown in the appendix the ${\bf k} \neq {\bf k}'$ sum can be safely neglected
due to random phase cancellation and we arrive at the final expression for
the full photoemission intensity expressed in Eq. \ref{eq:convBroad}.

Finally, we note that, in general, the linewidth ($\Gamma_m$) measured in ARPES from well
prepared, atomically flat surfaces is a function of the initial state or photohole linewidth ($\Gamma_i$) as well as the linewidth of the
final state or photoelectron ($\Gamma_f$). However, for the case of 2D states such as surface states in metals or thin
films such as graphene, there is no dispersion with $k_\perp$
and $\Gamma_m = \Gamma_i$.\cite{smith1993photoemission}

\subsection{\label{sec:Fitting}Analysis of Spectra and Discussion}

\begin{figure}
\includegraphics[width=8cm]{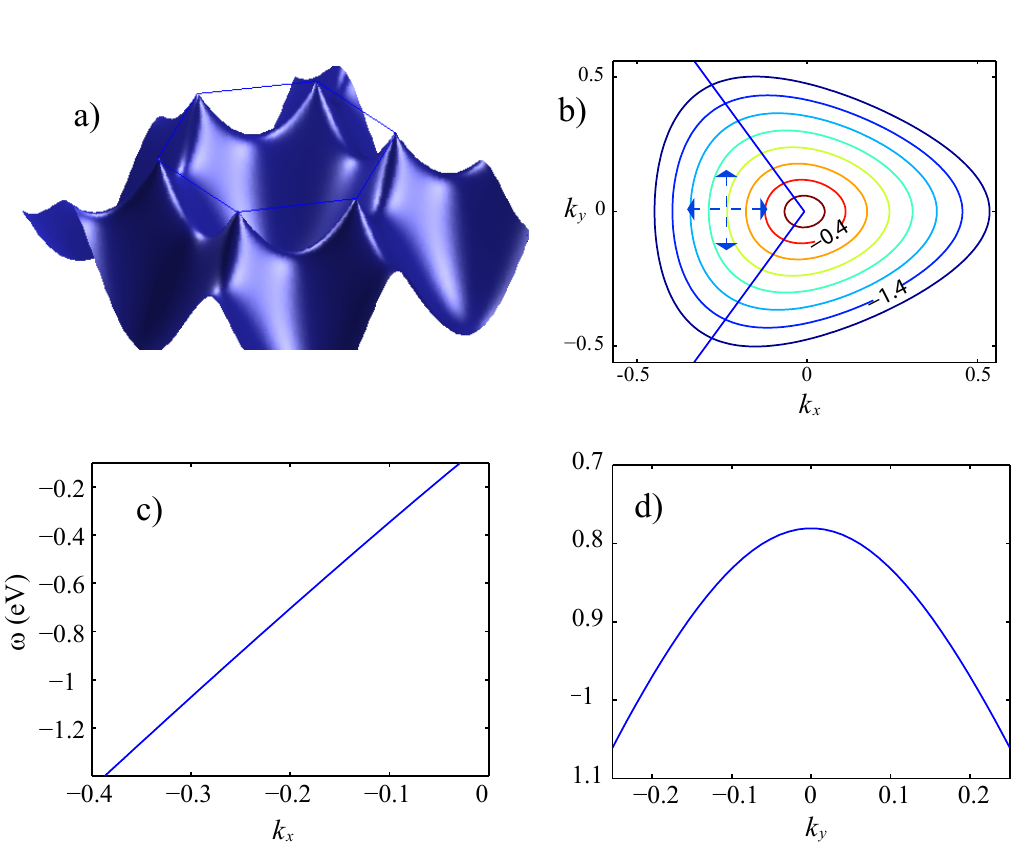}
\caption{\label{fig:bandstruct} (Color online) (a) Graphene bandstructure over first BZ.
(b) Contour lines drawn along constant binding energies in the vicinity of
the K point (binding energies indicated in eV). (c-d) Dispersion along $k_x$
and $k_y$, respectively (along dashed blue lines in (b)). }
\end{figure}

Figure \ref{fig:grapheneARPES} shows a plot of the dispersion obtained along
the $\Gamma$K direction in the vicinity of the Dirac point.
The average Fermi
velocity, derived from the slope of $\omega$ vs $k_x$ is $1.07\pm0.05 \times 10^6$ m/s.
This value is in excellent agreement with results obtained 
by IR measurements on undoped supported exfoliated graphene.\cite{jiang2007infrared,li2008dirac}
 Additionally, the dispersion along $\Gamma$K is linear with no deviations from linearity within our experimental uncertainty. As discussed above, despite
the roughness induced broadening in the spectrum, the
dispersion curve is easily extracted from the raw ARPES data by taking the second derivative of the ARPES intensity along the momentum direction. However, determining the intrinsic width of spectral features requires a deeper analysis.

Our prior measurements of the surface corrugation in suspended graphene allow
us to extract the intrinsic electronic structure from our ARPES data. 
The procedure for this fitting is as follows: first, $S_{k_\perp}$ is determined
from our surface morphology measurements and used as a constant parameter,
then
$A({\bf k_\|)}$ is varied until the convolved function, $I({\bf k_\|})$, represents a good
fit to the experimental data. Although a full deconvolution is, in principle,
possible it is much more straightforward to begin with an assumption for
the functional form of $A({\bf k_\|)}$ and systematically vary the parameters
until a good fit is found.   
The functional form of $A({\bf k_\|)}$ is assumed to be a Lorentzian, the
most commonly used photoemission lineshape, which results
from the $\bf k$-independent approximation for $\text{Im} [\Sigma({\bf k},\omega)]$.\footnote{
In fitting the MDCs
an instrumental broadening term was also convolved with our spectral function.
The energy resolution introduces a width of $\Delta E v_F$. In combination with the
lateral resolution of the instrument (known from prior calibration), this
results in a Gaussian response function with a width of approximately 0.042 $\text{\AA}^{-1}$.}

\begin{figure}
\includegraphics[width=8cm]{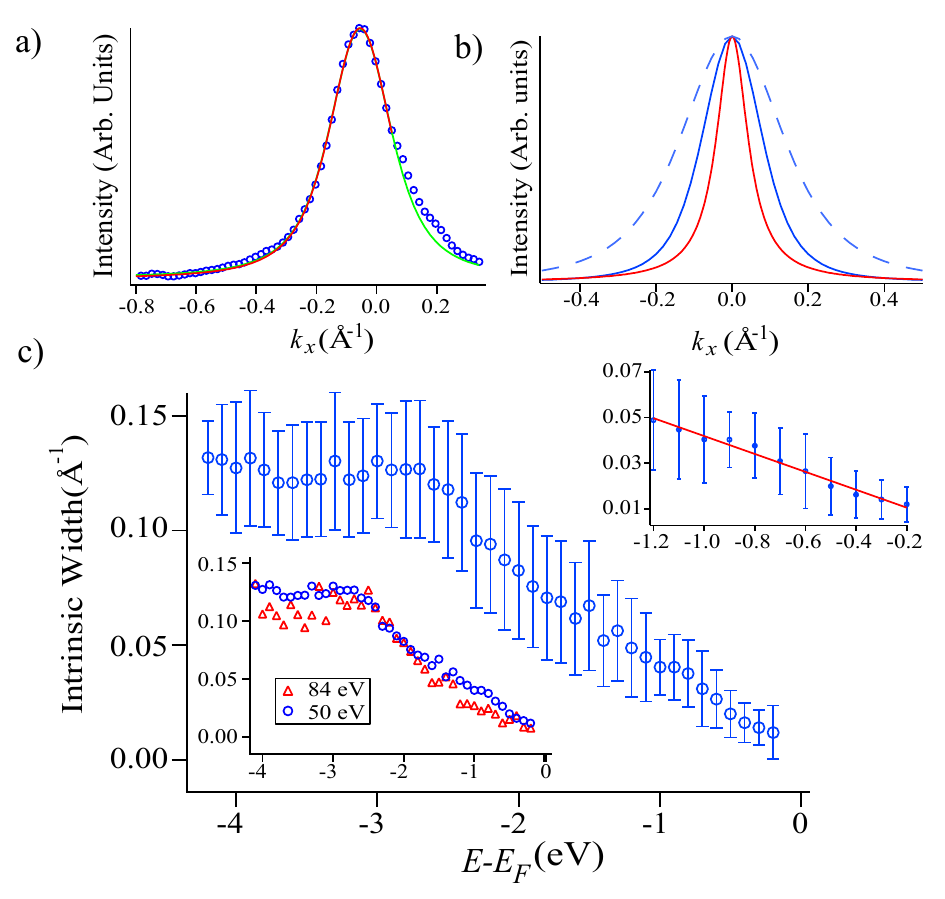}
\caption{\label{fig:selfEn} (Color online) Intrinsic width of ARPES features
for Suspended monolayer graphene. (a) Example of MDC fitting. (b) Two independent contributions
to broadening. Red line: intrinsic linewidth of initial state. Blue line: broadening due to corrugation
at $\hbar \omega$=50 eV (solid) and $\hbar \omega$=84 eV (dashed). (c) Inverse lifetime as a function
of binding energy for $\hbar \omega$=50 (blue) and $\hbar \omega$=84
eV (red). Inset shows best fit line to intrinsic width vs. binding energy
for $\hbar \omega$=50 data in vicinity
of Fermi level. }
\end{figure}

In carrying out this procedure, we introduce two additional simplifications.
First, note that we are examining the region in $\bf k$-space within the first Brillouin
zone along the $\Gamma$K ($k_x$) direction in the vicinity of the K point. As shown in Fig. \ref{fig:bandstruct}, the $\pi$ state disperses rapidly along $k_x$
in this region,
but relatively slowly along $k_y$ since $\partial \omega / \partial k_y$=0 at
$k_y$=0. Thus, although Eq. \ref{eq:mBroad} describes a 2D convolution, it is possible to replace the required 2D $k_x k_y$ integral with a 1D integral along $k_x$.
Second, we note that most of the MDCs considered here have an asymmetric peak shape, with
additional spectral weight in the $\omega>v_F |{\bf k}|$ region of the curve. Possible
reasons for this asymmetry are discussed in a separate paragraph below. For fitting purposes, this additional spectral weight was not considered and the best fit
was obtained by imposing a momentum cutoff within 0.1 $\text{\AA}^{-1}$ of the peak position on the $\omega>v_F |{\bf k}|$ side of the curve.
Figure \ref{fig:selfEn}(a) shows a representative
curve from the $\hbar \omega$=50 eV data taken 0.7 eV below the Fermi level
along with a best fit. Note that the lineshape of this curve provides
an excellent fit to the experimental data. 
Figure \ref{fig:selfEn}(b) shows
the two independent contributions to the linewidth: the corrugation-induced
broadening and the intrinsic broadening.  In order to cross-check that the convolution procedure accurately captures the photon-energy dependence of the photoemission process, the same fitting procedure
was repeated on data obtained with a photon energy of $\hbar \omega$=84. At
this photon energy, $k_\bot = 4.27 \text{\AA}^{-1}$ and, according to Eq. \ref{eq:mBroad}, the width of
$S_{k_\perp}$ is nearly twice as large as it is at $\hbar \omega = 50$ eV.
However, as expected, the intrinsic linewidth extracted from the fitting
procedure is the same for data obtained with both photon energies. A comparison of the self-energy
extracted from the two data sets is shown in Fig. \ref{fig:selfEn}(c); 
the two resulting curves are the same, within experimental error, thus confirming the
photon-energy dependence given in Eq. \ref{eq:mBroad} and lending further support
to our approach.

To make our observations quantitative and enable comparison with other work,
we perform a linear fit of the intrinsic width versus
binding energy, $\Gamma_i=\alpha + \beta (E-E_F)$. From this fit, we find
$\alpha=0.002 \pm 0.005$  $\text{\AA}^{-1}$ and $\beta=0.039 \pm 0.01$  $\text{\AA}^{-1} eV^{-1}$.
 As expected, the value of $\alpha$ is within experimental uncertainty
of zero since excited states just above the Fermi level should be very long
lived.
Consider now the parameter $\beta$ that describes the increase in inverse quasiparticle lifetime with increasing binding energy. 
The lifetime is related to $\Gamma_i$ by $\tau=1/(2 \Gamma_i
v_F)$.
Thus, our measured value can be reexpressed as $\beta$= 0.78$\pm$0.02 fs$^{-1}$ eV$^{-1}$, so as to enable ready comparison with prior measurements of the same quantity on graphite and exfoliated graphene.
For graphite, $\beta$ has been measured by femtosecond photoemission to be an order of magnitude smaller, viz. 0.029 fs$^{-1}$ eV$^{-1}$,\cite{xu1996energy} while STS measurements of exfoliated graphene on graphite have
produced an intermediate value of ($\beta$ = 0.11 fs$^{-1}$ eV$^{-1}$).\cite{li2009scanning} A reasonable explanation for this discrepancy is the greater out-of-plane corrugation of
suspended graphene, which has been predicted to be the largest contribution to electron scattering in rough graphene sheets\cite{gazit2009theory,mariani2008flexural,katsnelson2008electron}. Indeed, such roughness constitutes
short-range correlated disorder, which has also been shown theoretically to lead to scattering rates which scale linearly with $\omega$ in graphene.\cite{foster2008graphene}

Comparison can also be made with results obtained on epitaxial graphene grown
on SiC. In such a system the Dirac point is $\sim$ 0.5 eV below the Fermi
level which changes the quasiparticle dynamics resulting in a non-linear
behavior for $\Gamma_i$ vs $\omega$. In particular, it has been shown that
electron-plasmon interaction in doped epitaxial graphene results in an increase in the electron scattering
rate in a narrow energy region where $\omega \sim E_D$.\cite{bostwick2006quasiparticle}
However, at deeper binding energies a nearly linear increase of $\Gamma_i$
has been demonstrated with a slope of $\sim$ 0.025 $\text{\AA}^{-1} eV^{-1}$,
which is comparable to our measured value of $\beta$.

Because of the unique Dirac Fermion behavior and two-dimensionality of graphene, there has been much discussion
of many-body physics that would lead to lifetime broadening in
ARPES measurements of graphene\cite{bostwick2006quasiparticle,das2007many,gonzalez1996,gonzalez2001electron}. 
In conventional bulk crystals Fermi-liquid theory predicts the decay of a photohole through creation of an electron-hole pair to result in a lifetime which scales as $(E - E_F )^2$, in proportion to the number of excitation pathways that satisfy momentum and energy conservation. However, the linear dispersion of the graphene bands along with the vanishing density of states at $E_F$ modify this picture.
Hence, undoped graphene is expected to show anomalous marginal Fermi-liquid behavior, characterized by a lifetime that scales as $(E - E_F)^{-1}$.\cite{das2007many} Electron-phonon
interaction has also been shown experimentally to lead to linewidth broadening.\cite{li2008dirac,bostwick2006quasiparticle}
However, the interaction is limited by the phonon dispersion to within 140
meV of $E_F$.\cite{liu2010phonon} Coulombic interactions, however can affect
scattering rates for electrons
 well below $E_F$. 
As noted above, elastic scattering
due to short range correlated impurities such as adatoms, dislocations or corrugations has also been shown theoretically to produce a $(E-E_F)^{-1}$
dependence on lifetime.\cite{foster2008graphene} 

As discussed above, prior
STS measurements have confirmed this linear increase for a small range of
energies ($\sim$150 meV) in the vicinity of the Fermi level for exfoliated
graphene on graphite.\cite{li2009scanning} Our measurement confirms that
this behavior persists as far as 2eV below the Fermi level; a log-log plot
of $\Gamma_i$ vs $\omega$ displays a slope of $\sim$1. As noted above,
such marginal Fermi-liquid behavior has also been observed by femtosecond time-resolved
photoemission spectroscopy on graphite.\cite{xu1996energy}

We now return to the topic of asymmetry in MDC peak shape. 
As many recent theoretical studies have pointed out, the commonly made $\bf k$-independent approximation for $\text{Im} [\Sigma({\bf k},\omega)]$ is not fully valid in graphene as the doping level approaches zero.\cite{das2007many,gonzalez1996}
The vanishing density of states at $\omega=E_F$ along with graphene's linear dispersion near $E_F$ places a kinematic restriction on the available phase
space for electron-electron scattering. The scattering pathway $e^- \rightarrow
e^- + e^-h^+$ is only available for off-shell electrons for which $\omega>v_F |{\bf k}|$ and is kinematically forbidden when $\omega<v_F |{\bf k}|$. Thus, one expects
a discontinuity in $\text{Im} [\Sigma({\bf k},\omega)]$ at $\omega = k$
and decay due to electron-electron interaction
may be indicated by asymmetry in MDC peak shape.\cite{foster2008graphene,gonzalez1996}  As mentioned above and indicated
in Fig. \ref{fig:selfEn}, in MDCs taken through the K point for monolayer
graphene
additional spectral weight is present in the $\omega>v_F |{\bf k}|$ regime. In principle, a full deconvolution of the ARPES intensity would recover the
exact function form of $A({\bf k},\omega)$. However, such a procedure would require use of
the full 2D integral specified by Eq. \ref{eq:convBroad}, which is beyond
the scope of the work presented here.

\section{\label{sec:sum}Summary}

Photoemission on thin sheets of 2D crystals
is expected to grow in importance as interest in single layer insulators and semiconductors
increases. We have performed ARPES on a 2D suspended surface with well defined
surface corrugation. By comparing our work with our prior results obtained
from diffraction measurements on corrugation in suspended graphene sheets\cite{knox2008spectromicroscopy}
we have developed a model for understanding the effect of corrugation
on ARPES spectra. By analyzing results obtained with different photon excitation
energies, we have estimated the contribution of surface roughness
to broadening. Thus, despite the surface corrugation in the graphene layer,
it is still possible to develop insights into graphene physics. In particular,
we have shown that exfoliated suspended graphene is essentially undoped in
its pristine form. Additionally, we have shown that the band structure has
no significant deviations from linearity in the vicinity of the Dirac point.
Our measured Fermi velocity is
comparable to results
obtained on supported graphene by transport and optical measurements. Finally,
we have also shown that undoped exfoliated graphene behaves as
a marginal Fermi-liquid with an anomolous carrier lifetime, which scales
as $(E-E_F)^{-1}$.

\begin{acknowledgments}
K.R.K. acknowledges support for materials and sample preparation from the
NSF Number CHE-0641523 and by NYSTAR; R.M.O. and K.R.K. acknowledge support
from DOE BES  (Contract No. DE-FG02-04-ER-46157) for experimental work. 
K.R.K and R.M.O acknowledge support from NSF number 0937683 for travel.
The synchrotron portion of the project (A.M., D.C.) at Elettra was supported
also through PRIN2008-prot.20087NX9Y7\_002. A.M. gratefully acknowledges
the NSEC at Columbia University and the Italian Academy at Columbia University
for the warm hospitality and financial support during his visit. The authors
also thank Matthew Foster and Mark Hybertsen for several extensive and helpful discussions.
\end{acknowledgments}

\appendix*

\section{}

In this appendix, we will follow the standard formalism for single photon photoemission using the dipole approximation. We will
adapt the treatment to deal with a locally curved surface using a specific
initial state described by the tight binding model for graphene.

According to the standard tight-binding scheme, initial $\pi$ states in the
valence band of graphene with energy $\omega_k$ and crystal-momentum
${\bf k}$ are represented as a linear combination of molecular $p_z$ orbital
states:

\begin{equation}
  \psi_{\bf k} ({\bf r})=\frac{1}{\sqrt{N}} \sum_{\bf R} e^{i {\bf k} \cdot
  {\bf R}} \sum_{j=A,B} C^{\bf k}_j \phi ({\bf r}-{\bf R}-\tau_j)  
\end{equation}
where $\frac{1}{\sqrt{N}}$ is an overall normalization factor, $A$ and 
$B$ designate the sublattice sites and $\tau_j$ their locations within the
unit cell. The $C^{\bf k}_j$ are complex coefficients obtained from the tight-binding
model and the $\phi$ are molecular $p_z$ orbitals. The sum over ${\bf R}$ runs over all $N$ unit cells in the crystal (note that we work in the limit where $N\rightarrow\infty$).

The transition-matrix for photoexcitation from this initial state to a plane
wave final state with total momentum $\bf{k}_f$ outside the crystal can be
written, using  the dipole approximation, as follows:

\begin{equation}
  M^{\bf{k}} \propto \int d^3 r e^{-i {\bf k}_f \cdot {\bf r}} ({\bf p} \cdot
  {\bf A})
  \psi_{\bf{k}_i} ({\bf r}) 
\end{equation}

Inserting the above definition for the initial state we obtain:

\begin{equation}
  M^{\bf{k}} \propto ({\bf k_i} \cdot \hat{\bf{\lambda}}) 
  \sum_R e^{i ({\bf k_i}-{\bf k_f}) \cdot {\bf R}} \tilde{\phi}(k_f)
  \sum_{j=A,B} C^{\bf k}_j e^{-i{\bf k_f} \cdot \tau_j}
\label{eq:Tmatrix}
\end{equation}
where $\tilde{\phi}$ is the Fourier transform of the molecular $p_z$ orbital
and $\hat{\bf{\lambda}}$ represents the polarization vector of the incoming
radiation. We are interested in a small region of momentum-space in the vicinity
of the K point. Since ${\bf k_i} \cdot \hat{{\bf \lambda}}$ and $\tilde{\phi}({\bf
k_f})$ are nearly constant in this region, we concentrate our attention on
the sum, $\sum e^{i ({\bf k_i}-{\bf k_f}) \cdot {\bf R}} \sum C^{\bf k_i}_j e^{-i{\bf k_i} \cdot \tau_j}$. The sum over $j$, $\sum C^{\bf k_i}_j e^{-i{\bf k_f} \cdot
\tau_j}$, depends only on the relative phase between the $C^{\bf k}_j$'s and
the pathlength difference from atoms $A$ and $B$ to the detector. This term
changes rapidly on a contour around the K point. Along the $\Gamma$K direction,
the term changes from 2 to 0 as we pass through the K point from the first
to the second BZ. However, if we restrict ourselves to the region of ${\bf k}$-space
along the $\Gamma$K direction within the first BZ (see Fig \ref{fig:bandstruct}), $\sum C^{\bf k_i}_j e^{-i{\bf k_f} \cdot
\tau_j}$ is nearly constant. Thus, we are left with
the sum $\sum e^{i ({\bf k_i}-{\bf k_f}) \cdot {\bf R}}$.

For a 3D crystal with perfect translational symmetry ${\bf R}=n_1{\bf a_1}+n_2{\bf
a_2}+n_3{\bf a_3}$ where $n_1,n_2,n_3$ are integers and ${\bf a_1},{\bf a_2},{\bf
a_3}$ are primitive lattice vectors. In this case, the sum over ${\bf R}$
reduces to momentum preserving delta function $\delta({\bf k_f}-{\bf k_i} -{\bf
G})$ where $G$ is a reciprocal lattice vector. 
However, since graphene is a two-dimensional lattice,
momentum conservation does not hold in the perpendicular direction.
More significantly, exfoliated graphene is a flexible membrane that is not
atomically flat, so the {\bf R}'s must be expressed in terms of a continuous
variable; thus ${\bf R} = n_1 \textbf{a}_1+ n_2 \textbf{a}_2 + \Delta x + \Delta y + z$, where $z_j$ is a
continuous variable which represents the local height of the graphene sheet.
The variation in height is such that we can consider the well known theory
of scattering from continuous rough surfaces in order to evaluate the sum
in equation \ref{eq:Tmatrix}. We begin by replacing
the discrete sum with an integral:
 
\begin{equation}
  \sum_{\bf R} e^{i ({\bf k_i}-{\bf k_f}) \cdot {\bf R}} =  \int d^3 r D({\bf
  r}) e^{i ({\bf k_i}-{\bf k_f}) \cdot {\bf r}}
\end{equation}
where $D(r)$ is the density-function of the material which, for a perfectly
crystalline flat sample, has the form:

\begin{equation}
  D({\bf r}) = \sum_{\bf R} \delta ({\bf r} - {\bf R})
\end{equation}
 
A periodic lattice generates a photoemission spectrum
with the periodicity of the reciprocal lattice. However,
since we are concerned with the photoemission spectrum in a small region
of $\bf k$-space in the first Brillouin zone, we may
abandon the description of the surface as a discrete lattice and replace
it with a smooth, continuous sheet. Thus, we approximate $D(r)$ as
a surface density function:
 
\begin{equation}
  D({\bf \rho},z') \simeq \delta [z' - z({\bf \rho})]
\end{equation}
where $\mathbf{\rho}=r_{\|}=(x,y)$ and $z'=r_{\perp}$. Thus, the surface
is now defined by the height function $z=z(\mathbf{\rho})$.
Inserting the above definition of $D(r)$ and explicitly separating
$\bf k_i$ and $\bf k_f$ into components parallel and perpendicular to the surface
we obtain:

\begin{equation}
   M \propto \int d^2 \rho e^{i ({\bf k_i}_{\perp}-{\bf k_f}_{\perp}) \cdot z(\rho)}
  e^{i ({\bf k_i}_{\|}-{\bf k_f}_{\|}) \cdot \rho}
\end{equation}
We have retained momentum conservation; for a constant $z(\rho)$ the above
integral produces $\delta({\bf k_i}_{\|}-{\bf k_f}_{\|})$ times a complex phase; but for
a non-trivial $z(\rho)$, the delta function broadens since
$ e^{i ({\bf k_i}_{\perp}-{\bf k_f}_{\perp}) \cdot z(\rho)}$ is no longer
independent of $\rho$. Additionally, electronic states in graphene propagate
on a curved space, which implies that the direction of the initial state
wavevector, ${\bf k_i}$, varies as a function of position along the surface
so that $k_{i \perp}$ and $k_{i \|}$ vary with $\rho$ as well. This introduces
additional phase variation into the exponential argument $({\bf k_i}_{\perp}-{\bf
k_f}_{\perp}) \cdot z(\rho)$. However, this variation is very small in comparison
to that introduced by changes in $z(\rho)$ and can effectively be ignored
with little change in our final result. In particular, $k_{i \perp}$ varies proportional
to $k \frac{\partial z}{\partial \rho_k}$ which is on the order of 0.01 \AA$^{-1}$.
Thus the phase variation in the term $k_{i \perp} z \ll 1$ ($\Delta z \approx$
2 \AA) is very small in comparison to the variation in the $k_{f \perp} z$ term
($k_{f \perp}$ ranges from 2.5 to 3.5 \AA$^{-1}$). Thus, we will approximate
the direction of the initial state wavevector, ${\bf k_i}$, as constant for
all points on the surface. This means that $\bf k$ is not $\rho$ dependent and
we can define a new vector ${\bf k} = {\bf k_i} - {\bf k_f}$, so that our expression
becomes:

\begin{equation}
   M \propto \int d^2 \rho e^{i {\bf k}_{\perp} \cdot z(\rho)}
  e^{i {\bf k}_{\|} \cdot \rho}
\end{equation}

To find the photoemission intensity we use Fermi's golden rule which yields:

\begin{subequations}
\begin{eqnarray}
I \propto \frac{2 \pi}{\hbar} \|M\|^2   \nonumber
\\
\|M\|^2 \propto \int d^2 \rho' d^2 \rho e^{i {\bf k}_{\perp} \cdot z(\rho)}
\nonumber
\\
\times e^{i {\bf k}_{\|} \cdot \rho}e^{-i {\bf k}_{\perp} \cdot z(\rho')} e^{-i
{\bf k}_{\|} \cdot \rho'}
\end{eqnarray}
\end{subequations}

Defining $r=\rho-\rho'$ we can rearrange to obtain:

\begin{equation}
   I \propto \int d^2 r \left(\int d^2 \rho e^{i {\bf k}_{\perp}
   \cdot [z(\rho+r)-z(\rho)]}\right) e^{i {\bf k}_{\|} \cdot r}
\end{equation}

The term inside the parenthes is 
the height-difference function, $C(r,k_\perp)$, of the surface which 
is related to the height-height correlation function, $H(r)=\langle |z(r_0+r)-z(r_0)|^2
\rangle$.  
It is straight forward to show that the $C(r,k_\perp)$ equals 
$e^{-\frac{1}{2}H(r)k_\perp^2}$.\cite{yang1993diffraction}
Thus, we have:

\begin{subequations}
\begin{eqnarray}
I \propto \int d^2 r C(k_\perp,r) e^{i {\bf k}_{\|} \cdot r},
\\
C(k_\perp,r) = e^{-\frac{1}{2}H(r)k_\perp^2}.
\end{eqnarray}
\end{subequations}

For a large class of surfaces, $H(r)$ has the following properties:

\begin{eqnarray}
H(r) \propto e^{2\alpha}, \text{for } r\ll\xi
\\
H(r) = 2w^2, \text{for } r\gg\xi
\end{eqnarray}
where $\alpha$ is a measure of the small scale roughness termed the ``roughness
exponent." In, particular, it can be shown that the full width at half maximum
(FWHM)
of $I(k_{\|})$ (with $k_\perp$ held constant) scales as $\xi^{-1}
(w k_\perp)^{\frac{1}{\alpha}}$ when $(w k_\perp)^2\gg1$. The functional
form of $I(k_{\|})$ is well approximated as:\cite{yang1993diffraction}

\begin{eqnarray}
  S_{k_\perp}({\bf k_\|}) = (\xi / (w k_\perp)^{\nicefrac{1}{\alpha}})
  F_{\alpha}({\bf k_\|} \xi / (w k_\perp)^{\nicefrac{1}{\alpha}})
  \nonumber\\
  F_{\alpha}(Y)=\int X dX exp(-X^{2 \alpha} ) J_0 (XY)
\end{eqnarray}

The above discussion began with the assumption that the initial state, $\psi_k$,
had a well defined pseudo-momentum, $k$, and energy $\omega_k$. To include
initial state broadening in our description, we replace $\psi_k$ with a sum
over multiple momentum states, $\sum a_k \psi_k$, where the $a_k$ are complex
coefficients. The coefficients, $a_k$, are related to the spectral function,
$A(k,\omega)$ by $|a_k|^2=A(k,\omega)$ with the spectral function defined
as:

\begin{equation}
  A(k,\omega)=\frac{Im(\Sigma)}{(\omega_k-\omega-Re(\Sigma))^2+Im(\Sigma)^2}
\end{equation}
where $\Sigma=\Sigma(k,\omega)$ is the quasiparticle self-energy.  Retaining
our simple description of the final state as a free-electron state with momentum
$q$, our transition matrix becomes a sum, $M=\sum a_k M^{kq}$, over multiple
matrix elements weighted by the complex coefficients $a_k$, where the $M^{kq}$
are the original transition matrix elements defined in Eq. \ref{eq:Tmatrix}.
Again, using Fermi's golden rule we find that the transition probability
is proportional to the square of this sum:

\begin{eqnarray}
  I \propto |M|^2 = |\sum_k a_k M^{kq}|^2 = 
  \nonumber\\
  \sum_k |a_k|^2 |M^{kq}|^2 + \sum_{k
  \neq k'}a_k^* a_{k'} M^{kq*} M^{k'q}
\end{eqnarray}

The cross terms have the form:

\begin{equation}
   M^{kq*} M^{k'q} \propto \int d^2 r \left(\int d^2 \rho e^{i {\bf k}_{\perp}
   \cdot [z(\rho+r)-z(\rho)]}e^{i \Delta {\bf k}_{\|}\rho}\right) e^{i {\bf
   k'}_{\|} \cdot r}
\end{equation}
where $\Delta {\bf k}_{\|} = {\bf k'}_{\|} - {\bf k}_{\|}$, ${\bf k}_{\perp}
\approx {\bf k'}_{\perp}$. The $e^{i \Delta {\bf k}_{\|}\rho}$ factor in
the $\rho$ integral introduces a random phase that causes the integral to
average to zero (since it is taken over the whole surface). Thus, the cross
terms can be safely neglected and we arrive at the final expression or photoemission
intensity as a function of $k_\|$ with $k_\perp$ fixed, described by Eq. \ref{eq:convBroad}.

\bibliography{endnotes}

\end{document}